\documentstyle[aps,prl,twocolumn,psfig,floats]{revtex}
\begin{document}
\preprint{\vbox{\hbox{UCB-PTH-99/51}},
  \vbox{LBNL-44551}}
\draft
\wideabs{
\title{Neutrino Mass Anarchy}
\author{Lawrence Hall, Hitoshi Murayama and Neal Weiner}
\address{
Department of Physics, 
University of California, 
Berkeley, CA~~94720, USA;\\ 
Theory Group, 
Lawrence Berkeley National Laboratory, 
Berkeley, CA~~94720, USA}

\date{\today}
\maketitle


\setcounter{footnote}{0}
\setcounter{page}{1}
\setcounter{section}{0}
\setcounter{subsection}{0}
\setcounter{subsubsection}{0}

\begin{abstract}
What is the form of the neutrino mass matrix which governs the 
oscillations of the atmospheric and solar neutrinos? Features of the data
have led to a dominant viewpoint where the mass matrix has an ordered,
regulated pattern, perhaps dictated by a flavor symmetry. We challenge
this viewpoint, and demonstrate that the data are well accounted for by a
neutrino mass matrix which appears to have random entries.
\end{abstract}
}
\vskip 0.3in
\noindent {\bf 1} 
Neutrinos are the most poorly understood among known elementary
particles, and have important consequences in particle and nuclear
physics, astrophysics and cosmology. Special interests are devoted to
neutrino oscillations, which, if they exist, imply physics beyond the
standard model of particle physics, in particular neutrino masses.
The SuperKamiokande data on the angular dependence of 
the atmospheric neutrino flux provides strong evidence for neutrino 
oscillations, with $\nu_\mu$ disappearance via large, near maximal 
mixing, and $\Delta m^2_{atm} \approx 10^{-3}$ eV$^2$\cite{SKatm}. Several 
measurements of the solar neutrino flux can also be interpreted as 
neutrino oscillations, via $\nu_e$ disappearance\cite{SKsolar}. 
While a variety of 
$\Delta m^2_\odot$ and mixing angles fit the data, in most cases
$\Delta m^2_\odot$ is considerably lower than $\Delta m^2_{atm}$, and 
even in the case of the large angle MSW solution, the data typically 
require $\Delta m^2_\odot \approx 0.1 \Delta m^2_{atm}$\cite{valle}. The 
neutrino mass matrix apparently has an ordered, hierarchical form for the 
eigenvalues, even though it has a structure allowing large mixing angles. 

All attempts at explaining atmospheric and solar neutrino fluxes in terms 
of neutrino oscillations have resorted to some form of ordered, highly 
structured neutrino mass matrix\cite{models}. These 
structures take the form $M_0 + 
\epsilon M_1 + ...$, where the zeroth order mass  matrix, $M_0$, contains 
the largest non-zero entries, but has many zero entries, while the first 
order correction terms, $\epsilon M_1$, have their own definite texture, 
and are regulated in size by a small parameter $\epsilon$. 
Frequently the pattern of the zeroth 
order matrix is governed by a flavor symmetry, and the hierarchy of mass 
eigenvalues result from carefully-chosen, small, symmetry-breaking 
parameters, such as $\epsilon$. 
Such schemes are able to account for both a hierarchical 
pattern of eigenvalues, and order unity, sometimes maximal, mixing. Mass 
matrices have also been proposed where precise numerical ratios of different 
entries lead to the desired hierarchy and mixing.

In this letter we propose an alternative view. This new view selects
the large angle MSW solution of the solar neutrino
problem, which is preferred by the day to night time flux ratio at the
$2 \sigma$ level\cite{SKsolar}. While the masses and 
mixings of the charged fermions certainly imply regulated, hierarchical 
mass matrices, we find the necessity for an ordered structure in the neutrino 
sector to be less obvious. Large mixing angles would result from a 
random, structureless matrix, and such large angles could be responsible 
for solar as well as atmospheric oscillations. Furthermore, in this case 
the hierarchy of $\Delta m^2$ need only be an order of magnitude, much 
less extreme than for the charged fermions. We therefore propose that the 
underlying theory of nature has dynamics which produces a neutrino mass 
matrix which, from the viewpoint of the low energy effective theory, 
displays {\it anarchy}: all entries are comparable, no pattern or 
structure is easily discernable, and there are no special precise 
ratios between any entries. Certainly the form of this mass matrix is not 
governed by approximate flavor symmetries.

There are four simple arguments against such a proposal
\begin{itemize}
\item{The neutrino sector exhibits a hierarchy with
    $\Delta m^2_{\odot} \approx 10^{-5}-10^{-3}\mbox{eV}^2$ for the large 
    mixing angle solution, while $\Delta
    m^2_{atm} \approx 10^{-3}-10^{-2}\mbox{eV}^2$,}
\item{Reactor studies of $\overline\nu_e$ at the CHOOZ experiment have
    indicated that mixing of $\nu_e$ in the $10^{-3}\mbox{eV}^2$
    channel is small \cite{CHOOZ}, requiring at least one small angle,}
\item{Even though large mixing would typically be expected from anarchy, {\it
maximal} or near maximal mixing, as preferred by SuperKamiokande data, would be
unlikely,}
\item{$\nu_e, \nu_\mu$ and $\nu_\tau$ fall into doublets with
    $e_L$, $\mu_L$ and $\tau_L$, respectively, whose masses are
    extremely hierarchical ($m_e:m_\mu:m_\tau \approx
    10^{-4}:10^{-1}:1$).}
\end{itemize}
By studying a sample of randomly generated neutrino mass matrices, we 
demonstrate that each of these arguments is weak, and that, even when 
taken together, the possibility of neutrino mass anarchy still appears quite 
plausible.

\noindent {\bf 2}
We have performed an analysis of a sample of random neutrino matrices. 
We investigated three types of neutrino mass
matrices: Majorana, Dirac and seesaw. For the Majorana type, we
considered $3\times 3$ symmetric matrices with 6 uncorrelated parameters. For
the Dirac type, we considered $3 \times 3$ matrices with 9
uncorrelated parameters. Lastly, for the seesaw type, we considered
matrices of the form $M_D M_{RR}^{-1} M_D^T$\cite{seesaw}, 
where $M_{RR}$ is of the former type and $M_D$ is of the latter.
We ran one million sample matrices with independently generated
elements, each with a uniform distribution in the interval $[-1,1]$
for each matrix type: Dirac, Majorana and seesaw.

To check the 
robustness of the analysis, we ran smaller sets using a distribution 
with the logarithm base ten uniformly distributed in the interval 
$[-1/2,1/2]$ and with random sign. We further checked both of these 
distributions but with a phase uniformly distributed in $[0, 2 \pi]$.
Introducing a logarithmic distribution and phases did not significantly 
affect our results (within a factor of two), 
and hence we discuss only matrices with a linear distribution and real entries.

We make no claim that our distribution is somehow physical, nor do we 
make strong quantitative claims about the confidence intervals of
various parameters. However, if the basic prejudices against anarchy
fail in these simple distributions, we see no reason to cling to them.

\begin{figure}
  \centerline{
    \psfig{file=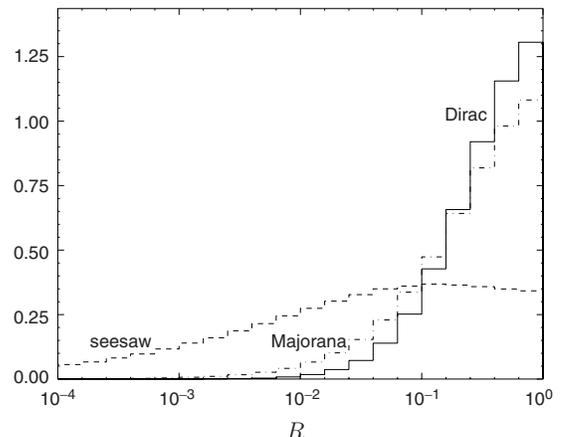,width=0.4\textwidth,angle=0} }
        \caption[sniff]{The distribution of $\Delta
          m^2_{\odot} / \Delta m^2_{atm}$ for Dirac (solid) Majorana
          (dot-dashed) and seesaw (dashed) scenarios.}
        \label{fig:massdistributions}
\end{figure}

In each case we generated a random neutrino mass matrix, which we 
diagonalized with a matrix $U$. We then investigated the following 
quantities:
\begin{eqnarray}
        R     &\equiv& \Delta m^{2}_{12} / \Delta m^{2}_{23}, \\
        s_{C} &\equiv& 4 |U_{e3}|^{2}(1-|U_{e3}|^{2}),  \\
        s_{atm} &\equiv& 4 |U_{\mu3}|^{2}(1-|U_{\mu3}|^{2}), \\
        s_{\odot} &\equiv& 4 |U_{e2}|^{2} |U_{e1}|^{2},
\end{eqnarray}
where $\Delta m^{2}_{12}$ is the smallest splitting and $\Delta 
m^{2}_{23}$ is the next largest splitting. 
What ranges of values for these parameters should we demand from our
matrices? We could require they lie within the
experimentally preferred region. However, as experiments improve and
these regions contract, the probability that a random matrix will
satisfy this goes to zero. Thus we are instead interested in mass
matrices that satisfy certain {\it qualitative} properties. For our
numerical study we select these properties by the specific cuts
\begin{itemize}
        \item{$R < 1/10$ to achieve a large hierarchy in the $\Delta
m^2$. } 
        \item{$s_{C} < 0.15$ to enforce small $\nu_{e}$ mixing through this 
        $\Delta m^{2}$.}
        \item{$s_{atm}>0.5$ for large atmospheric mixing.}
        \item{$s_{\odot}>0.5$ for large solar mixing.}
\end{itemize}
\noindent The results of subjecting our $10^6$ sample matrices, of Dirac,
Majorana and seesaw types, to all possible combinations of these cuts
is shown in Table \ref{tb:splits}.
\begin{table}
 \begin{center}
  \begin{tabular}{||l|r|r|r|r||}
Dirac~~~~~~~& no cuts & $s_{atm}$ & $s_{\odot}$ & $s_{atm}+s_{\odot}$ \cr
\hline
no cuts& 1,000,000& 671,701& 184,128& 135,782\cr
\hline
$s_{C}$&  145,000& 97,027& 66,311& 45,810\cr
\hline
$R$& 106,771& 78,303& 17,538 &14,269\cr
\hline
$s_{C}+R$& 12,077& 9,067& 5,656& 4,375
\end{tabular}
  \begin{tabular}{||l|r|r|r|r||}
Majorana& no cuts & $s_{atm}$ & $s_{\odot}$ & $s_{atm}+s_{\odot}$ \cr
\hline
no cuts & 1,000,000& 709,076& 200,987 &164,198\cr
\hline
$s_{C}$& 121,129 &91,269 &70,350& 56,391\cr
\hline
$R$& 200,452& 149,140& 37,238& 31,708\cr
\hline
$s_{C}+R$& 21,414 &16,507 &12,133 &10,027
\end{tabular}
  \begin{tabular}{||l|r|r|r|r||} 
seesaw~~~~\hskip0.1in  & no cuts & $s_{atm}$ & $s_{\odot}$ & $s_{atm}+s_{\odot}$ \cr
\hline
no cuts & 1,000,000& 594,823& 210,727 &133,800\cr
\hline
$s_{C}$& 186,684& 101,665& 86,511& 49,787\cr
\hline
$R$& 643,394& 390,043& 132,649& 86,302\cr
\hline
$s_{C}+R$& 115,614& 64,558& 53,430& 31,547
\end{tabular}
\end{center}
\caption{Mass matrices satisfying various sets of cuts for the real 
linear Dirac, Majorana and seesaw scenarios.}
\label{tb:splits}
\end{table}
First consider making a single cut. As expected, for all types of
matrices, a large percentage (from 18\% to 21\%) of the random
matrices pass the large mixing angle solar cut, and similarly for the
large mixing angle atmospheric cut (from 59\% to 71\%). Much more
surprising, and contrary to conventional wisdom, is the relatively
large percentage passing the individual cuts for $R$ (from 10\% to 64\%)
and for $s_C$ (from 12\% to 18\%).
The distribution for $R$ is shown in Figure \ref{fig:massdistributions}. 
Naively, one might expect that this would peak at $R=1$, which is 
largely the case for Dirac matrices, although with a wide peak. 
In the Majorana 
case there is an appreciable fraction $(\sim 20\% )$ that have a 
splitting $R\le 1/10$, while in the seesaw scenario the {\it majority} 
of cases $(\sim 64 \% )$ have a splitting $R\le 1/10$ 
--- it is not at all unusual to generate a large hierarchy.

 We can understand this simply: 
first a splitting of a factor of $10$ in the $\Delta m^{2}$'s 
corresponds to only a factor of $3$ in the masses themselves if they 
happen to be hierarchically arranged. Secondly, in the seesaw scenario, 
taking the product of three matrices spreads
the  $\Delta m^{2}$ distribution over a wide range.

\begin{figure}
  \centerline{ 
    \psfig{file=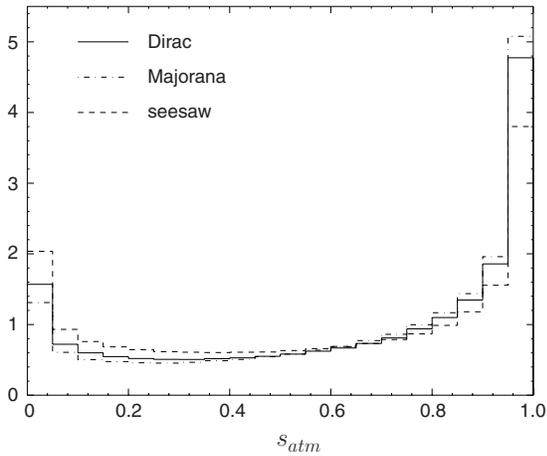,width=0.4\textwidth,angle=0}}
        \caption[sniff]{Plots of the normalized, binned distributions 
        of $s_{atm}$ for Dirac, Majorana and seesaw  
        cases. Contrary to intuition, the distributions actually peak 
        at large $s_{atm}$.}
        \label{fig:atmplots}
\end{figure}

\begin{figure*}
  \centerline{ \psfig{file=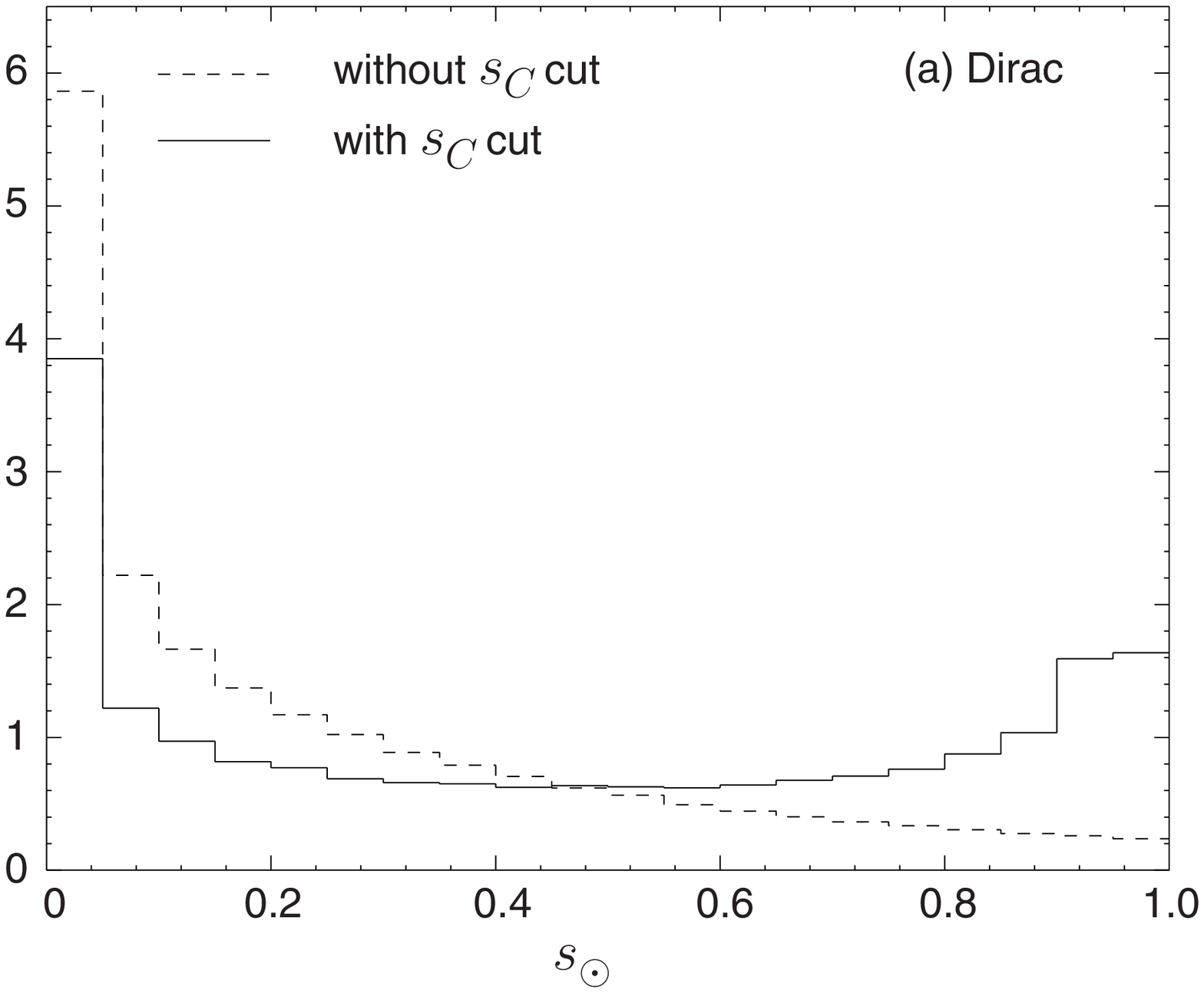
      ,width=0.3\textwidth,angle=0}
          \psfig{file=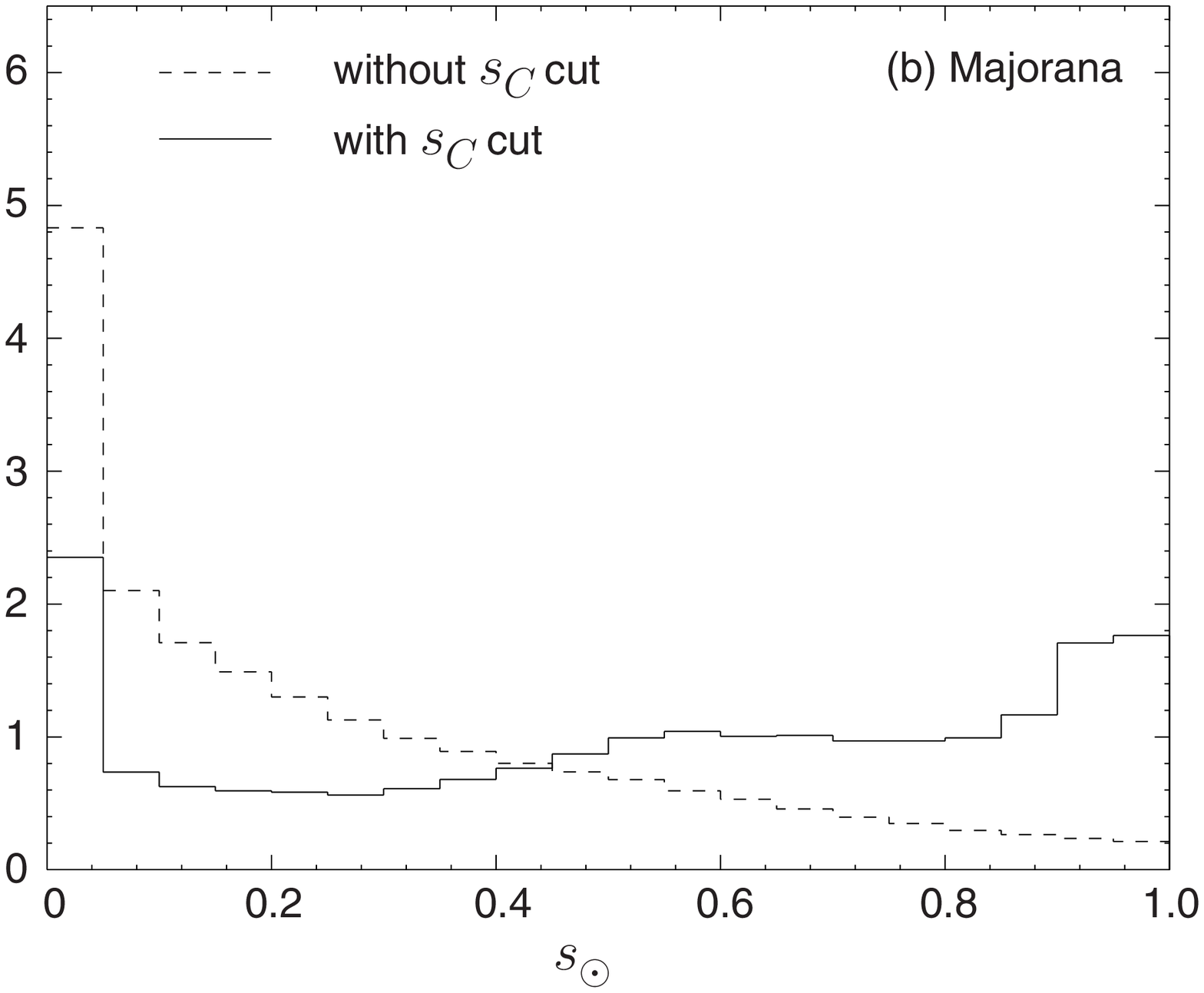,width=0.3\textwidth,angle=0}
\psfig{file=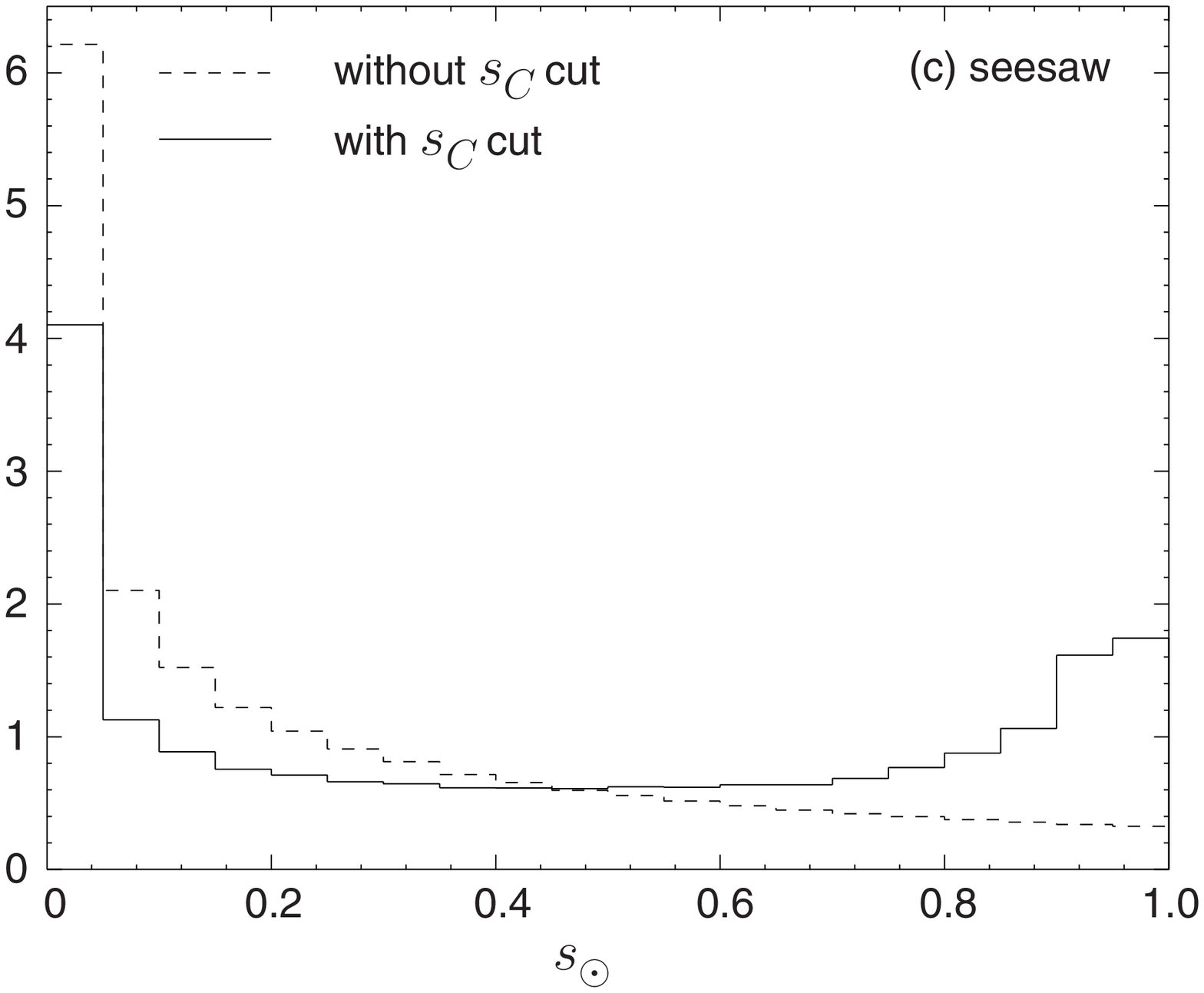,width=0.3\textwidth,angle=0} }
        \caption[sniff]{Plots of the normalized, binned distributions 
        of $s_{\odot}$ for Dirac (a), Majorana (b), and seesaw (c) 
        cases. The distribution after imposing the $s_C$ cut (solid) 
        shows a greater preference for large $s_{\odot}$ compared with 
        the original distribution (dashed).}
        \label{fig:solarplots}
\end{figure*}

While one would expect random matrices to typically give large
atmospheric mixing, is it plausible that they would give near-maximal
mixing, as required by the SuperKamiokande data?
In Figure \ref{fig:atmplots} we show distributions of 
$s_{atm}$, which actually {\it peak} in the $0.95 < s_{atm}<1.0$ bin. 
We conclude that it is not necessary to impose a
precise order on the mass matrix 
to achieve this near-maximal mixing.
Finally, we consider correlations between the various cuts. For example,
could it be that the cuts on $R$ and $s_C$ selectively pass matrices
which accidentally have a hierarchical structure, such that $s_{atm}$
and $s_{\odot}$ are also small in these cases? From Table 
\ref{tb:splits} we see that
there is little correlation of $s_{atm}$ with $s_C$ or $R$: the fraction of
matrices passing the $s_{atm}$ cut is relatively insensitive to
whether or not the $s_C$ or $R$ cuts have been applied. However, there
is an important anticorrelation between $s_{\odot}$ and $s_C$ cuts;
for example, in the seesaw case roughly half of the matrices 
satisfying the $s_C$ cut satisfy the $s_\odot$ cut, compared with $20\% $ 
of the original set. This anticorrelation is shown in more detail in Figure
\ref{fig:solarplots}, which illustrates how the $s_C$ cut serves to
produce a peak at large mixing angle in the $s_\odot$ distribution.

For random matrices we expect the quantity
\begin{equation}
       s_{C}+s_{\odot} =  
4 (|U_{e1}U_{e2}|^{2}+ |U_{e1}U_{e3}|^{2}+ |U_{e2}U_{e3}|^{2})
            \label{eq:totalmixing}
\end{equation}
to be large, since otherwise $\nu_e$ would have to be closely aligned 
with one of the mass eigenstates. Hence, when we select matrices where
$s_C$ happens to be small, we are selecting ones where $s_\odot$ is 
expected to be large. 

\noindent {\bf 3}
We have argued that the neutrino mass matrix may follow from complete
anarchy, however the electron, muon, tau mass hierarchies imply that
the charged fermion mass matrix has considerable order and
regularity. What is the origin for this difference? The only answer
which we find plausible is that the
lepton doublets, $(\nu_l, l)_L$, appear randomly in mass operators,
while the lepton singlets, $l_R$, appear in an orderly way, for
example, regulated by an approximate flavor symmetry. This
idea is particularly attractive in SU(5) grand unified theories where
only the 10-plets of matter feel the approximate flavor symmetry,
explaining why the mass hierarchy in the up quark sector is roughly
the square of that in the down quark and charged lepton sectors.  
Hence we consider a charged lepton mass matrix of the form 
\begin{equation}
M_l =  \hat{M}_l \pmatrix{\lambda_e&0&0 \cr 0& \lambda_\mu&0 \cr
0&0&\lambda_\tau} 
\end{equation}
where $\lambda_{e,\mu,\tau}$ are small flavor symmetry breaking
parameters of order the corresponding Yukawa couplings, while
$\hat{M}_l$ is a matrix with randomly generated entries. 
We generated one million neutrino mass matrices and 
one million lepton mass matrices, and provide results for the mixing
matrix $U= U_l^\dagger U_\nu$, where $U_\nu$ and $U_l$ are the unitary
transformations on $\nu_l$ and $l_l$ which diagonalize the neutrino 
and charged lepton mass matrices.
We find that the additional mixing from the charged leptons does not
substantially alter any of our conclusions -- this is illustrated for
the case of seesaw matrices in Table \ref{tb:su5splits}.
The mixing of charged leptons obviously cannot affect $R$, but 
it is surprising that the distributions for the mixings
$s_{atm,\odot,C}$ are not substantially changed.

\begin{table}[t]
 \begin{center}
  \begin{tabular}{||l|r|r|r|r||}
 cuts & none & $s_{atm}$ & $s_{\odot}$ & $s_{atm}+s_{\odot}$ \cr
\hline
none &  1,000,000& 537,936& 221,785& 126,914\cr
\hline
$s_C$& 222,389& 102,178 &99,050& 50,277\cr
\hline
$R$& 643,127& 345,427& 142,789& 81,511\cr
\hline
$s_C+R$& 143,713& 65,875& 63,988 &32,435
\end{tabular}
\end{center}
\caption{Mass matrices satisfying various sets of cuts for the real 
linear seesaw scenario, with additional mixing from the charged lepton 
sector.}
\label{tb:su5splits}
\end{table}
 
\noindent {\bf 4}
All neutrino mass matrices proposed for atmospheric and solar neutrino
oscillations have a highly ordered form. In contrast, we have proposed that the
mass matrix appears random, with all entries comparable in size and no
precise relations between entries. We have shown, especially in the
case of seesaw matrices, that not only are large mixing angles for
solar and atmospheric oscillations expected, but $\Delta m^2_{\odot}
\approx 0.1 \Delta m^2_{atm}$, giving an excellent match to the large angle
solar MSW oscillations, as preferred at the $2 \sigma$ level in the
day/night flux ratio. In a sample of a million random seesaw matrices,
40\% have such mass ratios and a large atmospheric mixing. Of these,
about 10\% also have large solar mixing while having small $\nu_e$
disappearance at reactor experiments. Random neutrino mass matrices
produce a narrow peak in atmospheric oscillations around the
observationally preferred case of maximal mixing. In contrast to
flavor symmetry models, there is no reason to expect $U_{e3}$ is
particularly small, and long baseline experiments which probe $\Delta
m^2_{atm}$, such as K2K and MINOS, will likely see large
signals in $\overline\nu_e$ appearance. If $\Delta m^2_{atm}$ is at the lower
edge of the current Superkamiokande limit, this could be seen at a
future extreme long baseline experiment with a muon source. 
Furthermore, in this scheme $\Delta m^2_\odot$ is large enough to be
probed at KamLAND, which will measure large $\bar{\nu}_e$ disappearance.

\begin{acknowledgements}
        This work was supported in part by the Director, Office of Science,
Office of High Energy and Nuclear Physics, Division of High Energy
Physics of the U.S. Department of Energy under Contract
DE-AC03-76SF00098 and in part by the National Science Foundation under
grant PHY-95-14797. 
\end{acknowledgements}



\end{document}